\documentclass[a4paper,11pt]{article}
\usepackage{latexsym}	
\usepackage{amsmath}
\usepackage{amsthm}

\usepackage{graphicx}
\def\ba{\begin{eqnarray}}
\def\ea{\end{eqnarray}}

\def\ba{\begin{eqnarray}}
\def\ea{\end{eqnarray}}

\def\lb{\label}
\def\be{\begin{equation}}
\def\ee{\end{equation}}

\theoremstyle{plain}

\begin{document}

\title{About gravitational radiation of semi local strings with non compact internal modes}
\author{Alejandro Morano\thanks{Departamento de F\'isica, UBA CONICET, Buenos Aires, Argentina, alejandro.d.morano@gmail.com} \; and Osvaldo P. Santill\'an\thanks{Instituto de Matem\'atica Luis Santal\'o (IMAS), UBA CONICET, Buenos Aires, Argentina,
		firenzecita@hotmail.com and osantil@dm.uba.ar.}}
	
	\maketitle
	
	\begin{abstract}
		The present work studies the gravitational radiation of a non abelian vortex with a non compact internal moduli describing its excitations. This situation may be realised by semi-local supersymmetric vortices \cite{novedad1}-\cite{reviewshifman}, although supersymmetry is not necessary for this to happen. In the situation considered along this work, the internal space has infinite volume, and a largely energetic perturbation propagates along the object, even though the vortex line may not be moving. A specific configuration is presented, in which the internal space is the resolved conifold with its Ricci flat metric. The curious feature about it is that it corresponds to a static vortex, that is, the perturbation is only due to the internal modes. Even being static, the emission of gravitational radiation is in the present case of considerable order. This suggest that the presence of slowly moving objects that can emit a large amount of gravitational radiation is a hint of non abelianity.
	\end{abstract}

	\section{Introduction}

	Gauge theories giving rise to non abelian vortices are of interest and not necessarily unrealistic. Such topological objects, even though not being present in the Standard Model at low temperature and low density phases, are likely to appear at its completions or in other of its phases with very high temperature or high densities. An example is the hypothetical colour-flavor locked phase of QCD \cite{cfl1}-\cite{cfl2}, which may be realized at the core of a neutron star. At such high densities, the colour superconductivity effect may play an important role in the dynamics. For QCD and other analogous models, the resulting non abelian vortices posses internal moduli being described by a compact space. An extensive pedagogical introduction about this subject can be found in \cite{nitta3} or \cite{2}, and we refer the reader to these references for details. 
	
	The gravitational and other channel radiations for non abelian vortices with compact moduli space were studied in \cite{yo}. The findings of that work show that the resulting radiation is similar to the one for abelian objects, but with a corrected loop factor. This factor depends on the shape of the vortex, and deviation from their expected values may be a hint of non abelianity.  It should be emphasized that, although the QCD vortices described above are an interesting example of topological objects with internal excitations, they posses an scale two low for generating a noticeable spectrum of gravitational waves. The results of \cite{yo} assume the presence of such type of vortices, but the symmetry breaking scale is assumed to be much larger than the one corresponding to the colour flavour locked phase. An example could be the GUT or even the Planck scale.

	The aim of the present work is to study gravitational radiation of vortices whose internal excitations parameterize a non compact space. As it will be argued along the text, this non compact property may enhance the deviation from the abelian radiation. Such vortices may lead to a noticeable gravitational spectrum. As in \cite{yo}, it will be assumed that these objects may appear in a completion of the Standard Model, at early universe stages, although the explicit form of this model will be not specified. The present work instead assumes that objects of such type may be present at early universe stages, which seems for the present authors a reasonable hypothesis,  and it is aimed to analyze their gravitational spectrum. 

	Recall that, very loosely speaking, in a non abelian theory, there is a vacuum configuration invariant under the action of given group $G$, which represents the unbroken symmetry group  \cite{nitta3}-\cite{2}. 
	A typical vortex solution tends to different vacuums in different directions at the asymptotic region. However,  the fields $\Phi_i$ of the theory take non vacuum values near the core of the object. This core may be assumed as an small radius infinite cylinder aligned along the z axis, by simplicity. Typically, the action by elements of $G$ usually maps one vortex solution into another. In general, it is just
	a subgroup $H$ of $G$ which is enough for parameterizing the different vortex solutions. Under a small perturbation, the vortex may move or vibrate, and this define the translational modes. 
	On the other hand,  along a perturbed vortex, the field configuration varies with the $z$ position in time, and different field configurations are related by each other by an 
	action $h(z, t)$ of elements of $H$. Therefore, a perturbation of the vortex is described by elements of $H$ which depends on the vortex position and the time. These are the internal excitation modes and in favorable situations, $H$
is described in terms of a manifold.  
	
	For abelian vortices, it is well known that when the perturbation lengths are considerably smaller that the typical vortex size, the Manton approximation takes place \cite{manton}. 
	The only possible perturbations are translational ones, described by the Nambu-Goto action in four dimensions \cite{abrikosov}-\cite{nielsen}, see also \cite{anderson1}-\cite{anderson3}. Classically, this is equivalent to a Polyakov action. The possible emission channels of these objects were considered for instance in \cite{theisen}-\cite{peloso}. If the underlying gauge theory is non abelian instead, the excitations are customarily described by a Polyakov action whose metric has larger dimensions, due to the presence of excitations living on subgroup $H$, the internal modes \cite{nitta3}.
	
	The non abelian string to be considered below has a moduli space given by a product of the Minkwoski space, representing the translational modes, with a resolved conifold with its Ricci flat metric. The last part represents the internal modes. This non compact moduli space appears for instance in the $N=2$ supersymmetric context,
	for semi local strings \cite{novedad1}-\cite{reviewshifman}, see also, \cite{esfuerzo1}-\cite{fidel}. In those models, several extra flavors are added to a theory for which the moduli space is compact. This addition result in the emergence of size moduli, rendering the internal space non compact. The supersymmetry arguments fix the metric to be Ricci flat after taking into account non perturbative
	effects. 
	
	The choice of the Ricci flat conifold metric  as the internal manifold, which is employed in the present work,  is not particularly motivated by supersymmetry however. The reason for this choice is  more practical, and is that the authors were able to find an explicit string solution in
	which the object is static but nonetheless  the internal modes are evolving non trivially in time.  This leads to a non vanishing energy momentum tensor resulting in the emission of gravitational radiation. Therefore, although the object looks static, there appears a considerable gravitational wave emission, which may sound paradoxical. Such a pattern is pointing out that the vortex in consideration is  non abelian.  The goal of the present work is to exemplify this affirmation with an explicit and detailed calculation, and the conifold metric is suited for these purposes.
	
	The vortex excitations will be described by the Manton approximation \cite{manton}, that is, it will be assumed that the wave lengths of the excitations are smaller than the typical size of the object, as it will be specified along the text.

	The effective action for the motion for the translational and internal  modes will be taken as \cite{2}
	\begin{equation}\label{action}
		S=T_s\int_{\Sigma} \sqrt{-\eta}\eta^{ab}g_{\mu\nu} \partial_a x^\mu  \partial_b x^\nu d\Sigma,
	\end{equation}
	where $T_s$ is the string tension. This tension will not play a role in the derivation of the equations of motion. However it will be relevant when studying the gravitational waves the vortex emit when part of those modes are excited. The metric $g_{\mu\nu}$ is the direct sum of the Minkowski four dimensional metric with the conifold Ricci flat metric. The internal metric is multiplied by an internal tension $T_i$ which is not necessary  equal to the translational one $T_s$. However, as the internal space is non compact, by a simple scaling of the internal
	fields,  the action can be cast as in \eqref{action}.
	The vortex excitations depend on one spatial direction describing the vortex position  and one time direction
	describing the evolution. This determines a two dimensional surface $\Sigma$, that is, the world sheet metric, over which the last action integral is to be performed. The conformal gauge $\eta_{ab}=(-1, 1)$
	will be employed below, which leads to the standard Virasoro conformal constraints.
	
	The next sections are devoted to find configurations describing purely internal excitations and to characterize their gravitational radiation spectrum. The section 2 gives a light review about semi local vortices while section 3 is devoted to the integration of a particular solution. Section 4 presents
	 the calculation of the radiated gravitational power the object generates. Section 5 contains a discussion of the main results and their physical significance.
	
		\section{Semi local vortices as Polyakov strings in large dimensions}
	
	Before presenting particular results, it may be convenient to give a brief description about semi local vortices and their possible perturbations. Probably the most simple type of semilocal strings arise from the following action
	\be\lb{akshon}
	S=\int [\nabla_\mu \phi^i \nabla^\mu \overline{\phi}^{i}+\frac{g^2}{8}(\phi_i\overline{\phi}_i-v)^2+\frac{1}{4g^2}F_{\mu\nu} F^{\mu\nu}] d^4x,
	\ee
	where $i=1,2$ denote the complex scalar fields $\phi^i$ flavors, and there is only one gauge field $A_\mu$ involved, which is assumed to be abelian. In the previous formula, the covariant derivative 
	$$
	\nabla_\mu \phi_i=\partial_\mu\phi_i+\frac{i}{2} A_\mu \phi_i,
	$$
	has been introduced. If the parameter $v>0$ then the resulting photon is massive, with mass $m_\gamma\sim gv$. The minimum energy or vacuum condition is given by
	$$
	\phi_i\overline{\phi}_i=v. 
	$$
	This equation alone reduces the number of independent vacuum field from four to three. Still, there is a gauge transformation at hand which renders one of the fields, say $\phi_2$, real. This results in the algebraic condition for the vacuum manifold
	\be\lb{vaciamos}
	\phi_1\overline{\phi}_1+(\phi_2)^2=v,
	\ee
	which obviously defines the two sphere $S^2$, parametrizing the different vacuum orientations. 
	
	If in addition, the field $\phi^2$ is turned off, that is, $\phi^2=0$, then the resulting model is the abelian one studied by Abrikosov-Nielsen-Olesen (ANO) in the context of superconductivity \cite{ano}. The $S^2$ vacuum sphere now reduces to a circle. The model with the field $\phi^2$ turned off admits abelian vortex lines, the so called ANO vortices. If these vortices are aligned along the z axis, their profile is given by in terms of the transverse cylindrical coordinates $r$ and $\theta$ by
	\be\lb{ano}
	\phi_1=f(r) e^{i\theta},\qquad A_m=-\frac{2\epsilon_{mn}x_n}{r}(1-g(r)),
	\ee
	where the indices $m, n=x, y$ denotes the spatial cartesian directions transverse to the vortex. At large transverse coordinates $r\to \infty$ the corresponding fields tend to their vacuum values. Therefore the radial functions $g(r)\to 0$ and $f(r)\to v$, the first condition indicates that the gauge field is a pure gauge configuration. The scalar field winds the infinite region due to the factor $e^{i\theta}$. Near the origin $r\to 0$ the fields vanish, therefore $g(r)\to 1$ and $f(r)\to 0$. The precise form of the radial functions described above is given by the standard Bogomolnyi equations, but it is not know explicitly.
	
	Consider now small perturbation of the ANO vortex described above. A generic perturbation affects the position where the fields composing the vortex vanish. Obviously, if the vortex is unperturbed, this region is simply given by $x=y=0$, with $z$ and $t$ arbitrary. This is a two dimensional surface in the Minkowski four dimensional space time. For a perturbed vortex instead, the generalization of this region is given by $X^\mu(\tau, \sigma)$, which describes a moving line or string in three dimensional spatial directions or a two dimensional hypersurface in four dimensions. For the unperturbed vortex, it is clear that the identification $z=\sigma$ and $t=\tau$ takes place. If the vortex is perturbed instead, this line is deformed and varies with $z$ and $t$. 
	
	The equation of motions of the object can be schematically be derived as follows. First, the tangent space $T_p$ at a given space time point $p$ on the vortex may be parameterized by a vielbein $e_a=(t_i, p_j)$, with two components $t_i$ tangent to the surface $X^\mu(\tau, \sigma)$, and two remaining $p_j$ orthogonal to it. Any point close to the vortex may be then expressed as
	$$
	x_\mu=X_\mu(\tau, \sigma)+\eta^j p_j^\mu(\tau, \sigma).
	$$
	This defines a new coordinate system $(\tau, \sigma, \eta_1, \eta_2)$. If the curvature radius of the perturbation is small enough, the perturbed vortex fields composing the vortex may be obtained by the replacement $\phi(x,y)\to \phi(\eta_1, \eta_2)$ and $A^\mu(x, y)\to A^j(\eta_1, \eta_2)p_j^\mu(\tau, \sigma)$. After introducing this new profiles into the action \eqref{akshon} and by integrating along the transversal coordinates $\eta^i$, the resulting action becomes the Nambu-Goto one \cite{rubakov}
	\be\lb{ng}
	S_{ng}=T_s\int_{\Sigma} \sqrt{(\dot{X}\cdot X´)-(\dot{X})^2 (X)^2} d\Sigma.
	\ee
	The jacobian arises from the change of coordinates from $(x, y, z, t)$ to $(\tau, \sigma, \eta_1, \eta_2)$, and the tension $T_s$ arise due to the field integrations, and it is an indicator of how large are the fields composing the vortex and their derivatives along the three dimensional space where the vortex is placed. The last  action is equivalent, at classical level, to the Polyakov one
	\be\lb{pol}
	S_{p}=T_s\int_{\Sigma} \sqrt{-\eta}\eta^{ab}g_{\mu\nu} \partial_a X^\mu  \partial_b X^\nu d\Sigma,
	\ee
	with $g_{\mu\nu}$ denoting the standard Minkowski metric.
	The equations of motion corresponding to this action describe the possible perturbation profiles $X^\mu(\tau, \sigma)$ of the object. These excitations are refereed as the translational modes of the object. The range of validity of these equations is limited to perturbations which do not curve the object considerably. 
	
	The fact that the perturbation of the ANO abelian vortices are the translational ones is well known. On the other hand, as stated above, the ANO vortices are not the unique type of solutions that  \eqref{akshon} admits. A generalization of the above solution includes the scalar field $\phi_2$ turned on. The solution has the form \eqref{ano} but now includes the new real valued field \cite{semilocal2}
	$$
	\phi_2=h(r).
	$$
	The full solution tends to the vacuum if $h(r)\to 0$ at the asymptotic region $r\to \infty$. At the origin however, this field does not vanish, a feature that distinguishes these vortices from the ANO ones. The standard Bogomolnyi equation throws the following result
	\be\lb{anos}
	\phi_2=\frac{\rho}{r}f(r),
	\ee
	where $\rho$ is a complex parameter. This parameter is not constrained, and by reasons that will be clear below, it is known as a size parameter. Its presence leads to a long range power fall off at infinite, as the remaining profile functions $g(r)$, $f(r)$ defining the vortex decay as $\frac{r}{r^2+\rho^2}$ asymptotically. This type of objects  are known as semi local vortices, see \cite{semilocal2} and \cite{achu}.
	
	Consider now a perturbed vortex when the parameter $\rho$ is turned on, as in \eqref{anos}. On general grounds, it is expected that this parameter, which is constant for an unperturbed vortex, is prompted to a function $\rho(\tau, \sigma)$ for a perturbed one. In other words, the long range parameter varies along the vortex. The action \eqref{pol} has to be enlarged in order to include this perturbation. In addition, the vortex solution \eqref{anos} is not the only possible one, as it is evident that the interchange $\phi_1\leftrightarrow \phi_2$ leads to another valid vortex configuration. Instead on discussing the particular topological objects \eqref{anos} it is perhaps convenient at this point to focus on a more general  class of vortices,  not necessarily abelian, which are derived from a generalization of the action \eqref{akshon}, in which the gauge field $A_\mu$ and the scalar fields $\phi_j$ with $i=1,..., N$ belong to representations of certain non abelian group $G$. The potential of the theory $V(\phi_i)$ leads to vacuum conditions generalizing the ones in \eqref{vaciamos}, generically written as
	\be\lb{vaciamos2}
	Q(\phi_i)=0,
	\ee
	with $Q(\phi_i)$ a function derived from $V(\phi_i)$.
	The solution of this equation describe the different vacuum of the model. However, there may be a subgroup $H$ of $G$ which leaves the vacuum solutions  $\phi_{iv}$ unchanged. In general, these theories may admit vortex solutions generalizing \eqref{anos}, and may contain several size parameters $\rho_j$. The action of the group $H$ that leaves the vacuum invariant however, does not leave the vortex solution invariant, and maps different non equivalent vortex configurations while leaving their asymptotic behavior unchanged. In general, the group mapping different configurations may be smaller than $H$.  In any case, when the vortex is excited, it is expected that, as an observer moves along the object, he sees  translational modes $X^\mu(\tau, \sigma)$, together with size modes $\rho(\tau, \sigma)$ varying along the object. In addition, the vortex internal orientation changes as the observer moves, and these orientations are parameterized by elements $h(\tau, \sigma)$ of aforementioned subgroup of $H$.
	This last group may be compact, and in this  case the corresponding excitations are known as orientational modes. By assuming that these perturbations $\rho(\tau, \sigma)$ and $h(\tau, \sigma)$ are small, and by following a procedure analogous to the ones leading to \eqref{pol} it is found that the full spectrum of small excitations is described by the Polyakov action
	\be\lb{pol2}
	S_{p}=T_s\int_{\Sigma} \sqrt{-\eta}\eta^{ab}g_{\mu\nu} \partial_a X^\mu  \partial_b X^\nu d\Sigma,
	\ee
	where the metric $g_{\mu\nu}$ is the direct sum of the Minkowski one and an internal part $g_{int}$, generically non compact, and depending both on the size modes $\rho_i$ and the angular variables describing the elements $h$ of  $H$ which maps different orientational directions of the vortex. In all these considerations, the perturbations are assumed to be small. For instance, the difference $|\rho(\tau, \sigma)-\rho_0|<<\rho_0$ and so on. This is the generalization of the Manton approximation for the internal modes.
	
	In conclusion, the dynamics of non abelian vortex is in several contexts described by a Polyakov string action in dimensions larger than four, the extra dimensions describing the internal vortex excitations. The next task is to present examples of this situation.

	\section{Solutions of the equations of motion}
	
	As stated in the introduction, one of the tasks of the present work is to give examples of solutions of the action \eqref{action} in which no translational modes are excited. In other words, only the internal modes, size and orientational, are excited on the object.
	This will be exemplified with the resolved conifold metric, although the reader is encouraged to consider other non compact geometries. There are several references devoted to the conifold geometry and, in the following,
	the coordinates defined in \cite{3} will be employed.
	The local distance element is given as the direct sum of the Minkowski flat space and the resolved conifold metric, the local form is given by \cite{3}
	\begin{equation}\label{Vortex_metric}
		ds^2_{10} = -dt^2+dx^2+dy^2+dz^2+\kappa^{-1}(\rho)d\rho^2 + \frac{1}{9}\kappa(\rho)\rho^2 e_\psi^2 +\frac{1}{6}\rho^2(e_{\theta_1}^2+e_{\phi_1}^2)+ \frac{1}{6}(\rho^2+6a_c^2)(e_{\theta_2}^2+e_{\phi_2}^2),
	\end{equation}
	where the following radial function
	\begin{equation}\label{kappa_rho}
		\kappa(\rho)=\frac{\rho^2+9a_c^2}{\rho^2+6a_c^2},
	\end{equation}
	and the 6-bein
	\begin{equation*}
		e_\psi = d\psi + \sum_{i=1}^2 \cos{\theta_i} d\phi_i,\quad e_{\theta_i} = d\theta_i, \quad e_{\phi_i} = \sin\theta_i d\phi_i,\quad i=1,2,
	\end{equation*}
	were introduced. The particular solution of the equation of motion that will be found here corresponds to $\theta_1 = 0$, $\theta_2=\pi/2$, $\phi_1= \text{cte}$, $\phi_2=\phi$ and $x=y=0$. In this way, one is left with only one of the SU(2) parts of the conifold \cite{3} and the object will be static, standing straight along the z axis. The reduced metric becomes
	\begin{equation}
		ds^2_r = -dt^2+dz^2+\kappa^{-1}(\rho)d\rho^2 +\frac{1}{9}\kappa(\rho)\rho^2 d\psi^2 + \frac{1}{6}(\rho^2+6a_c^2)d\phi^2.
	\end{equation}
	In the terms given above, the effective lagrangian describing the motion of the vortex that follows from the Polyakov action \eqref{pol2} in the conformal gauge $\eta_{ab}=(-1, 1)$ is given by
	\begin{equation}\label{Lagrangian}
		\mathcal{L}=-(\dot{t}^2 - t'^2) + (\dot{z}^2 - z'^2) + \gamma(\rho)(\dot{\rho}^2 - \rho'^2) + \alpha(\rho) (\dot{\phi}^2 - \phi'^2) + \beta(\rho)(\dot{\psi}^2 - \psi'^2),
	\end{equation}
	where the following functions of $\rho$
	\begin{equation}\label{alpha_beta}
		\alpha= \frac{1}{6}(\rho^2+6a_c^2), \qquad
		\beta = \frac{1}{9}\kappa(\rho)\rho^2\qquad
		\gamma= \kappa^{-1}(\rho),
	\end{equation}
	have been introduced by simplicity.  In principle, there is an internal tension $T_i$ which may differ with respect to the $T_s$
	tension for the translational modes. Nevertheless, this can be adsorbed by a scaling $\rho^2\to T_s T^{-1}_i\rho^2$ and $a_c^2\to T_s T_i^{-1}a_c^2$, and the action will be as above, up to a constant factor $T_s$.

	The resulting equations of motion derived from this lagrangian are
	$$
	\alpha_\rho(\dot{\rho}\dot{\phi}-\rho'\phi') + \alpha(\Ddot{\phi}-\phi'')=0,
			$$
			$$
			\beta_\rho(\dot{\rho}\dot{\psi}-\rho'\psi')+\beta(\Ddot{\psi}-\psi'')=0
			$$
			$$
			\gamma_\rho(\dot{\rho}^2-\rho'^2)+2\gamma(\Ddot{\rho}-\rho'') = \alpha_\rho (\dot{\phi}^2-\phi'^2)+\beta_\rho(\dot{\psi}^2-\psi'^2),
		$$
	where the subscript $\rho$ here means partial derivative of this coordinate.
	
	For the purpose of getting a static vortex in $M_4$ with only excitations living at the conifold moduli, the following ansatz \cite{1}
	\begin{equation}\label{ansatz}
			t=\tau  \qquad
			z =\sigma,\qquad
			\rho = \rho(y), \qquad 
			\phi = \omega_1 \tau + f(y),\qquad
			\psi = \omega_2 \tau + g(y),
	\end{equation}
	will be employed, where the variable $y=b\tau + a\sigma$ has been introduced. This means that the functions $\rho(y)$, $f(y)$
	and $g(y)$ represent a wave propagating along the object, with speed $v=\frac{b}{a}$.
	
	After substitution of the last ansatz into the equations of motion for $\phi$ and $\psi$, the result is a total derivative with respect to $y$, whose integration leads to
	\begin{equation}\label{PHI_PSI_Solve}
			\alpha[ b\omega_1 + (b^2- a^2) f_y ] = A,\qquad
			\beta[ b\omega_2 + (b^2- a^2) g_y] = B,
	\end{equation}
	where $A$ and $B$ are the integration constants. Furthermore, by substituting the ansatz in the equation for $r$ and by taking the equations above into account, it is arrived to a single variable differential equation for $\rho(y)$
	\begin{equation}\label{Deq}
		(b^2-a^2)[ 2 \gamma \frac{\partial^2 \rho}{\partial y^2} + \gamma_\rho (\frac{\partial \rho}{\partial y})^2 ] = \alpha_\rho[ b^2(\frac{\omega_1}{b} + f_y )^2 - a^2 f_y^2 ] + \beta_\rho [ b^2(\frac{\omega_2}{b} + g_y )^2- a^2 g_y^2 ].
	\end{equation}
	The last expression can be cast in the following form
	\begin{equation}\label{Deq}
		(b^2-a^2)2\gamma^{1/2}\frac{\partial}{\partial y}[\gamma^{1/2}\;\frac{\partial \rho}{\partial y}] = \alpha_\rho[ b^2(\frac{\omega_1}{b} + f_y )^2 - a^2 f_y^2 ] + \beta_\rho [ b^2(\frac{\omega_2}{b} + g_y )^2- a^2 g_y^2 ].
	\end{equation}
	This is not the end of the story. The use of the conformal gauge $\eta_{ab}=(-1, 1)$ leads to the standard Virasoro constraints
	\begin{equation}\label{Virasoro}
			T_{\tau\tau} + T_{\sigma\sigma} = -\frac{2}{a^2+b^2} + \gamma \rho_y^2 + \alpha[f_y^2 + \frac{\omega_1^2+2b\omega_1 f_y}{a^2+b^2}]+ \beta[g_y^2 + \frac{\omega_2^2+2b\omega_2 g_y}{a^2+b^2}]=0, \end{equation}
			$$
			T_{\sigma\tau} = \gamma ab \rho_y^2 + \alpha(\omega_1+bf_y)af_y + \beta(\omega_2+bg_y)ag_y=0.
		$$
	The difference between these two Virasoro constraints gives the relation
	\begin{equation}
		-\frac{2}{a^2+b^2} + \alpha [-\frac{\omega_1}{b}f_y+\frac{\omega_1^2+2b\omega_1 f_y}{a^2+b^2}]+\beta [-\frac{\omega_2}{b}g_y+\frac{\omega_2^2+2b\omega_2 g_y}{a^2+b^2}]=0,
	\end{equation} 
	which can be expressed as well as follows
	\begin{equation}
		-2+ \frac{\omega_1}{b} \alpha[b\omega_1+(b^2-a^2)f_y]+\frac{\omega_2}{b}  \beta[b\omega_2+(b^2-a^2)g_y]=0.
	\end{equation}
	By using the equation (\ref{PHI_PSI_Solve}) it is deduced from the last formula the following relation between the constants involving the solution
	of the problem
	\begin{equation}\label{virasoro_constraint}
		\omega_1 A  + \omega_2 B=2b.
	\end{equation}
	This has to be supplemented with one of the equations (\ref{Virasoro}). The second of these equations can be considered as a conservation energy equation for a particle in one dimension, whose equation of motion in (\ref{Deq}) represent the Newton like equation with effective potential.  To see this fact clearly, it is convenient to introduce the potential
	\begin{equation}\label{efectos}
			V_{eff}(\rho)= \alpha\bigg(\frac{\omega_1}{b}+f_y\bigg)f_y + \beta\bigg(\frac{\omega_2}{b}+g_y\bigg)g_y
			=\frac{1}{(b^2-a^2)^2}[\frac{A^2}{\alpha}+\frac{B^2}{\beta}+\mu_1 \alpha+\mu_2\beta-\lambda ],
	\end{equation}
		with the constants
		\begin{equation}\label{def_lamda_mu}
			\lambda = 2 (b^2+a^2), \qquad \mu_i = a^2\omega_i^2.
		\end{equation}
		By employing the equations (\ref{PHI_PSI_Solve}),  it follows that (\ref{Deq}) and  (\ref{Virasoro}) can be written in the form
		\begin{equation}\label{Newton_curved}
				2\gamma^{1/2}(\rho)\frac{\partial}{\partial y}[\gamma^{1/2}(\rho)\;\frac{\partial \rho}{\partial y}] = - \frac{\partial V_{eff}}{\partial \rho}(\rho).\end{equation}
			which may be integrated in order to obtain
				\begin{equation}\label{integraste}
				\gamma(\rho) (\frac{\partial \rho}{\partial y})^2 + V_{eff}(\rho) = 0.
			\end{equation}
		The first represents a Newton law equation in presence of a potential, while the second results in the energy conservation corresponding to this equation, as stated above. 
		The last represents the energy conservation of a particle with variable mass $\gamma(\rho)$. This may sound a bit odd, since it is likely that in a variable mass system the  energy may be non conserved. In any case, the last equation is a first order integral of the previous one, and this is helpful for explicit integration. In addition, it is not difficult to see that the variation of the mass with respect to the "time" $y$ is not very pronounced, and this may simplify the task of finding its solutions. It may be helpful to write explicitly the last equation in the following form
		$$
		\bigg(\frac{\rho^2+6a_c^2}{\rho^2+9a_c^2}\bigg) \bigg(\frac{d\rho}{dy}\bigg)^2		
		+\frac{1}{(b^2-a^2)^2}\bigg[\frac{6 A^2}{\rho^2+6a_c^2}+\frac{9B^2}{\rho^2}\bigg(\frac{\rho^2+9a_c^2}{\rho^2+6a_c^2}\bigg)
		$$
		\be\lb{aintegrar}
		+\frac{\mu_1(\rho^2+6a_c^2)}{6}+\frac{\mu_2\rho^2}{9}\bigg(\frac{\rho^2+6a_c^2}{\rho^2+9a_c^2}\bigg)\bigg]=\frac{\lambda}{(b^2-a^2)^2}.
		\ee
		The variable mass term interpolates between the values $\gamma\sim 1/3$ and $\gamma\sim 1$, the first value is valid form small radius $\rho$ and the last for larger ones. 
		For $\rho<<\sqrt{6}a_c$ the equation results
		$$
		\frac{1}{3}\bigg(\frac{\partial \rho}{\partial y}\bigg)^2 + \widetilde{V}_{eff}(\rho) = 2E,
	    $$
		with a potential
		\begin{equation}\label{Veff_aprox}
			\widetilde{V}_{eff}(\rho) \simeq \frac{L^2}{ \rho^2 }+K \rho^2,
		\end{equation}
		with constants
		\begin{equation}\label{bilou}
			L^2 = 9 B^2, \qquad
			K = \frac{1}{4}(\mu_1+\mu_2-\frac{A^2}{6a_c^4}), \qquad
			E = \frac{3}{2}\lambda - \frac{A^2}{4a_c^2} - \frac{3}{2}a_c^2\mu_1,
		\end{equation}
		For the case $\rho >> 3a_c$ instead, the equation becomes
		$$
		\bigg(\frac{\partial \rho}{\partial y}\bigg)^2 + \widetilde{V}_{eff}(\rho) = 2E,
	    $$
	    where the effective potential has again the form \eqref{Veff_aprox} but with the new constants
		\begin{equation}\label{bilou2}
			L^2 = 6A^2+9 B^2, \qquad
			K = \frac{\mu_1}{6}+\frac{\mu_2}{9}, \qquad
			E = \frac{1}{2}\lambda.
		\end{equation}.
		By writing the effective interaction in the form (\ref{Veff_aprox}), it is seen that it corresponds to the radial part of a two dimensional harmonic oscillator. The parameter $L$ represents the angular momentum barrier, and $K$ is basically the oscillator constant, which is assumed to be equal in both directions. In other words, the oscillator is assumed to exert a central force on the particle with mass $m\sim \gamma$.
		
		The solution of a two dimensional central oscillator in Classical Mechanics is trivial. The lagrangian is simply 
		$$
		{\cal L}=\frac{m}{2}(\dot{x}^2+\dot{y}^2)-\frac{k}{2}(x^2+y^2),
		$$
		and it is clear that the equations of motion for $x$ and $y$ decouple. There are four integration constants that can be fixed with the initial positions $x_0$, $y_o$ and the velocities
		$\dot{x}_0$, $\dot{y}_0$.	
		The solution is elementary in Classical Mechanics. By passing to polar coordinates after solving these equations, it follows that the radial solution  can be written  as
		\begin{equation}\label{General_Sol2}2
			\rho^2 = \rho_0^2 \cos^2\Omega y + \bigg(\frac{E}{k}-\frac{\rho_0^2}{2}\bigg)\sin^2\Omega y + 2\rho_0 \dot{\rho}_0 \sin{2\Omega y},
		\end{equation}
		 $\dot{\rho}_0$ being the derivative of $\rho$ with respect to $y$ at $y=0$. The energy is given in terms of the radial initial conditions as
		\begin{equation}
			m \dot{\rho}_0^2 + \frac{l^2}{m\rho_0^2 }+k\rho_0^2=2E,
		\end{equation}
		while $l$ is fixed by the initial angular velocity $\dot{\theta}_0$ and radius $\rho_0$. By analogy, it follows that the solutions for small and large $\rho$ described above in \eqref{Veff_aprox} is given by
		\begin{equation}\label{General_Sol}2
			\rho^2 = \rho_0^2 \cos^2\Omega y + \bigg(\frac{E}{k}-\frac{\rho_0^2}{2}\bigg)\sin^2\Omega y + 2\rho_0 \dot{\rho}_0 \sin{2\Omega y},
		\end{equation}
		where now $\rho_0$ and $\dot{rho}_0$ are not independent since they must satisfy 
		\begin{equation}\label{relas}
			\gamma \dot{\rho}_0^2 + \frac{L^2}{\rho_0^2 }+K\rho_0^2=2E,
		\end{equation}
		and the parameters $L$, $K$ and $E$ are given in terms of the parameters of the model $A$, $B$ and $\mu_i$ and $\omega_i$ with $i=1,2$ described in \eqref{bilou}-\eqref{bilou2}.
		
		At this point it is convenient to recapitulate and write the full solution found along the text. First, the formula \eqref{ansatz} shows that the solution is given by
		\begin{equation}\label{ansatz2}
			t=\tau  \qquad
			z =\sigma,\qquad
			\rho = \rho(y), \qquad 
			\phi = \omega_1 \tau + f(y),\qquad
			\psi = \omega_2 \tau + g(y),
	\end{equation}
	  where now it has been determined that the function $\rho(y)$ is given by \eqref{General_Sol} together with \eqref{relas}. The constants $E$, $L$
	  and $K$ are given by \eqref{bilou} or \eqref{bilou2} for small and large $\rho$ excitations, respectively. These constants are given in terms of the parameters $A$, $B$,
	  and $\mu_i$ defined in \eqref{def_lamda_mu} and \eqref{PHI_PSI_Solve}, together with the condition which follows from the Virasoro constraints \eqref{virasoro_constraint} namely
	  \begin{equation}\label{def_lamda_mu2}
			\lambda = 2 (b^2+a^2), \qquad \mu_i = a^2\omega_i^2,
		\end{equation} 
		\begin{equation}\label{virasoro_constraint2}
		\omega_1 A  + \omega_2 B=2b.
	\end{equation}
In addition, recall from \eqref{PHI_PSI_Solve} that the constants $A$ and $B$ define the profile functions $f(y)$ and $g(y)$ by the first order equations of motion
		\begin{equation}\label{PHI_PSI_Solve2}
			f_y  =\frac{1}{b^2-a^2} \bigg[\frac{A}{\alpha}-b\omega_1\bigg],\qquad
			g_y = \frac{1}{b^2-a^2} \bigg[\frac{B}{\beta}b\omega_2\bigg],
	\end{equation}
	Recall that $\alpha(\rho)$ and $\beta(\rho)$ are functions of the radial coordinate $\rho$, whose explicit form are given in \eqref{alpha_beta}. Since for large and small radius the $\rho$  solution may be written as
		\begin{equation}\label{General_Sol3}
			\rho(y)^2 = U^2 \cos^2\Omega y + V^2\sin^2\Omega y + W \sin{2\Omega y},
		\end{equation}
		with the values of $U$, $V$ and $W$ follows by comparing the last expression with \eqref{General_Sol2}, it is seen that the first of equations \eqref{PHI_PSI_Solve2} is given by
		    $$
				f_y  =\frac{1}{b^2-a^2} \bigg[\frac{6A}{ U^2 \cos^2\Omega y + V^2\sin^2\Omega y + W \sin{2\Omega y}+6a_c^2}-b\omega_1\bigg].
		$$
		The second one is given by
				$$
				g_y = \frac{1}{b^2-a^2} \bigg[\frac{9B}{\rho^2}\bigg(\frac{\rho^2+6a_c^2}{\rho^2+9a_c^2}\bigg)-b\omega_2\bigg],
			$$
	 and the replacement of \eqref{General_Sol3} leads to a more complicated form. They can be integrated however, with the result
	 $$
	 (b^2-a^2)f(y) = \frac{6A}{\Omega\sqrt{6a^2+U^2}\sqrt{6a^2+V^2}} \arctan\left(\frac{\sqrt{6a^2+V^2}}{\sqrt{6a^2+U^2}}\tan\left(y\right)\right) - b\omega_1\;y, 
	 $$
	 $$
	 (b^2-a^2)g(y) = \frac{9\gamma B}{\Omega UV} \arctan\left(\frac{V}{U}\tan\left(y\right)\right) - b\omega_2\;y, 
	 $$
	 where the initial condition $f(0)=g(0)=W=0$ was chosen for simplicity. 
	 By substituting this solution into the ansatz (\ref{ansatz}),  the explicit form of the vortex solution
	 $$t=\tau, \quad z=\sigma, \quad y= b\tau +a\sigma $$
	 $$\rho = (U^2 \cos^2\Omega y + V^2\sin^2\Omega y + W \sin{2\Omega y})^{1/2}$$
	 $$\phi = \omega_1 \tau + \frac{6A}{\Omega (b^2-a^2)\sqrt{6a^2+U^2}\sqrt{6a^2+V^2}} \arctan(\frac{\sqrt{6a^2+V^2}}{\sqrt{6a^2+U^2}}\tan y)) - \frac{b\omega_1}{b^2-a^2}\;y$$
	 \be\lb{explicito}\psi = \omega_2 \tau + \frac{9\gamma B}{\Omega (b^2-a^2)UV} \arctan(\frac{V}{U}\tan y) - \frac{b\omega_2}{(b^2-a^2)}\;y.
	 \ee
	 is obtained. These solutions are valid for $\rho<<\sqrt{6}a_c$ or $\rho>>3 a_c$, but it will assumed that in the regime $\sqrt{6}a_c<\rho< 3 z_c$ the resulting solution, which is more complicated to be found analytically, have moderate values interpolating between these two regimes.

	\section{Gravitationally radiated power from the object}
	
	Having successfully derived a solution for the motion \eqref{explicito}, representing a non moving vortex with only internal excitations, the subsequent endeavor involves the investigation of the gravitational radiation emitted by the object. The rate of gravitational radiation emission at a frequency $\omega_n$ is described in the references \cite{weinberg}, \cite{vilenkin} by the formulas
	\begin{equation}\label{Power_solid}
		\frac{d P_n}{d\Omega} = \frac{G\omega_n^2}{\pi}[ T^*_{\mu\nu}(\omega_n,k)T^{\mu\nu}(\omega_n,k)-\frac{1}{2}|T^{\nu}_{\nu}(\omega_n,k)|^2].
	\end{equation}
	Here, $\frac{dP_n}{d\Omega}$ is the intensity of radiation at angular frequency $\omega_n=\frac{2\pi n}{T}$ per unit of solid angle in the direction of $k^i$, $k^ik_i = \omega_n^2$ and 
	\begin{equation}\label{T_transform}
		T^{\mu\nu}(k,\omega_n) =\frac{1}{T} \int_{-T}^{T} dt \; e^{i\omega_n t} \int_{R^3} d^3x\;  e^{-i k^i x_i}\;  T^{\mu\nu}(x^\mu)
	\end{equation}
	is the Fourier transform of the string energy-momentum tensor. As the preceding solution pertains to a very large straight string, the temporal periodicity $T$ is also a very large scale.
	
	The solutions that have been found in the previous section corresponds to a non moving string with only internal degrees of freedom excited. This internal motion however, gives rise to a non vanishing energy momentum tensor, therefore generating gravitational radiation. The task is now to determine whether this radiation is negligible or instead, if it takes values similar to a moving abelian vortex. If the second case is realized, then this radiation excess may be a hint for detecting non abelian objects.
	
	As is widely acknowledged, the energy-momentum tensor in the spatial coordinates $T^{\mu\nu}(x^\mu)$, can be obtained by varying the action of the string
	\begin{equation}\label{def_Tmunu}
		T^{\mu\nu}(x^\mu)=-\frac{2}{\sqrt{-g}}\frac{\delta (\sqrt{-g} \mathcal{L})}{\delta g^{\mu\nu}}.
	\end{equation}
	Here $g_{\alpha\beta}$ is the space time metric, assumed to be the flat Minkowski one. In these terms, the action describing the perturbations of the vortex \cite{2} that correspond to Lagrangian (\ref{Lagrangian}) in the conformal gauge becomes
	\begin{equation}
		S = T_s\int d^2\sigma\; \sqrt{-\gamma} \; \mathcal{L}  = T_s\int d^2\sigma\; \sqrt{-\gamma} \; M_{\alpha\beta}\;\partial_aX^\alpha \partial_bX^\beta.
	\end{equation}
	Here, $T_s$ represents the string tension, while $\gamma$ denotes the determinant of the metric on the worldsheet. In this scenario, the induced metric $\gamma_{ab}$ on the surface of the string is obtained as the pull-back of the complete metric of the vortex
	\begin{equation}
		\gamma_{ab} = g_{\alpha\beta}\;\partial_aX^\alpha \partial_bX^\beta.
	\end{equation}
	The metric $g_{\alpha\beta}$ describes all the string excitations, its explicit form being provided by the expression \eqref{Vortex_metric}, where its indices naturally range from $\alpha,\beta = 0,\dots,9$. Moreover, the fields $X^\alpha$ constitute the field excitations of the vortex, encompassing its coordinates in spacetime. In the current context, they correspond to those defined in equations \eqref{explicito}, that is $\rho$, $\phi$, $\psi$, $t$ and $z$. 
	
	The expression \eqref{Vortex_metric} reveals that the complete metric governing the string excitations is composed of two distinct components namely, the metric corresponding to translational modes, represented by a flat Minkowski metric $\widetilde{g}_{\mu\nu}=\eta_{\mu\nu}$, and the one describing the internal modes, represented by the Ricci-flat metric on the resolved conifold. Additionally, $\widetilde{g}=\eta=-1$. It is crucial to emphasize that even though $g_{\mu\nu}$ and $\widetilde{g}_{\mu\nu}$ are both given by the Minkowski 4-metric, they represent different quantities. One describes the internal modes, while the other characterizes spacetime where the vortex lives. The fact that they are both expressed in terms of the Minkowski metric is merely a coincidence, stemming from the assumption that the vortex exists in a flat spacetime.
	
	Hence, considering the earlier definition of the energy-momentum tensor and recognizing that the sole quantity dependent on the spacetime metric $g_{\mu\nu}$ is $\gamma$,  it is arrived to the ensuing formula
	\begin{equation}
		T^{\mu\nu}(x^\mu) = T_s \int d^2\sigma\; \frac{\mathcal{L}}{\sqrt{-\gamma}}\frac{\delta \gamma}{g_{\mu\nu}} \delta^4(x^\mu - X^\mu(\sigma,\tau)).
	\end{equation}
	Through the use of the Jacobi formula it is deduced that
	\begin{equation}
		\frac{\delta \gamma}{\delta g_{\mu\nu}} = \gamma\gamma^{ab} \frac{\delta \gamma_{ab}}{\delta g_{\mu\nu}} = \gamma\gamma^{ab} \partial_aX^\mu \partial_bX^\nu.
	\end{equation}
	By operating with the last three formulas it is straightforward to obtain the following expression for the stress-energy tensor of the configuration
	\begin{equation}
		T^{\mu\nu}(x^\mu) = T_s \int d^2\sigma\;  \mathcal{L}(\rho) \; \sqrt{-\gamma} \gamma^{ab} \partial_aX^\mu \partial_b X^\nu  \; \delta^4(x^\mu - X^\mu(\sigma,\tau)).
	\end{equation}	
	A remarkably simple expres)sion for $\mathcal{L}(\rho)$  can be derived as follows. First, by taking into account \eqref{Lagrangian} and \eqref{ansatz}, namely 
	\begin{equation}
		\mathcal{L}=-(\dot{t}^2 - t'^2) + (\dot{z}^2 - z'^2) + \gamma(\rho)(\dot{\rho}^2 - \rho'^2) + \alpha(\rho) (\dot{\phi}^2 - \phi'^2) + \beta(\rho)(\dot{\psi}^2 - \psi'^2),
	\end{equation}
	and
	\begin{equation}
		t=\tau  \qquad
		z =\sigma,\qquad
		\rho = \rho(y), \qquad 
		\phi = \omega_1 \tau + f(y),\qquad
		\psi = \omega_2 \tau + g(y),
	\end{equation}
	both considered in the previous section, it is obtained that
	$$
	{\cal L}=-2+(b^2-a^2)\gamma \rho_y^2+\alpha [(\omega_1+ b f_y)^2-a^2 f^2_y]+\beta [(\omega_2+ b g_y)^2-a^2 g^2_y].
	$$
	Now, the second term may be worked out by taking into account the Newton like equation \eqref{integraste} together with the effective potential \eqref{efectos}-\eqref{def_lamda_mu}. The remaining ones can be worked out with \eqref{PHI_PSI_Solve}, which allows to express the derivatives  $f_y$ and $g_y$ in terms of $\alpha(\rho)$ and $\beta(\rho)$. After some algebra, it is direct to find
		$$
	{\cal L}=-2-(b^2-a^2)V_{eff}(\rho)+\frac{A^2-\alpha^2 a^2\omega_1^2}{\alpha(b^2-a^2)}+\frac{B^2-\beta^2 a^2\omega_2^2}{\beta(b^2-a^2)}.
	$$
	By replacing the explicit expression for $V_{eff}$ given in \eqref{efectos}-\eqref{def_lamda_mu}, it is seen the terms proportional to $A^2$ and $B^2$ cancel and the final result becomes
	\begin{equation}
		\mathcal{L}(\rho) = -\frac{2a^2}{b^2-a^2}[\omega_1^2\alpha(\rho)+\omega_2^2\beta(\rho)-2].
	\end{equation}
	On the other hand, the metric of the worldsheet is
	\begin{equation}
		\gamma_{ab}=\text{diag}(-1, 1), \qquad  \sqrt{-\gamma} = 1.
	\end{equation}
	With the above derived formulas in hand, the next task is to study the vortex radiation.
	In a realistic situation, a vortex is large but finite. In order to apply \eqref{Power_solid} in a situation in which the vortex dimensions are large but finite, it is instructive to consider a static loop solution
	\begin{equation}
		x=\frac{L}{2\pi}\cos\sigma,\qquad
		y=\frac{L}{2\pi}\sin\sigma. \qquad
		t=\tau.
	\end{equation}
	This corresponds to a vortex entirely contained in the plane $z=0$ with circular form, and characteristic length $L$. In principle, the vortex tension will tend  to contract the object, and the loop will not be static after some time. However, if the vortex dimensions are large, it will not move for certain time and the most important contribution will be due to the internal motion. Furthermore, if the loop is extremely large it can be approximated by an straight configuration, such as the one described in the previous section. By considering all these approximations, the corresponding energy momentum tensor has the following non trivial components
	$$
	T^{\mu\nu} = T_s \int d^2\sigma\; \;\mathcal{L}(\rho)\;(\kappa_\tau^2 \dot{X}^\mu \dot{X}^\nu-\kappa_\sigma^2 X'^\mu  X'^\nu  )  \; \delta^4(x^\mu - X^\mu(\sigma,\tau)),
	$$
	$$ 
	T^{tt} = T_s \int d^2\sigma\;  \; \mathcal{L}(\rho)  \; \delta^4(x^\mu - X^\mu(\sigma,\tau)),
	$$
	$$
	T^{xx} = -T_s \int d^2\sigma\; \;\cos^2\sigma \mathcal{L}(\rho)  \; \delta^4(x^\mu - X^\mu(\sigma,\tau)),
	$$
	$$
	T^{yy} = -T_s \int d^2\sigma\; \sin^2\sigma \; \mathcal{L}(\rho)  \; \delta^4(x^\mu - X^\mu(\sigma,\tau)),
	$$
	\begin{equation}\label{T_components}
		T^{xy} = T_s \int d^2\sigma\; \;\sin\sigma\cos\sigma\; \mathcal{L}(\rho)  \; \delta^4(x^\mu - X^\mu(\sigma,\tau)).
	\end{equation}
	By introducing the quantities
	$$
	\omega = \frac{2\pi n}{T}=\frac{4\pi n}{L},\qquad k_x = \omega \alpha,\quad k_y = \omega \beta,\quad k_z = \omega \gamma,\quad \alpha^2+\beta^2+\gamma^2 = 1,
	$$
	together with the notation
	$$\Omega a  = \frac{2\pi}{L},\qquad \Omega b  =\frac{ 2\pi}{T} = \frac{ 4\pi}{L},$$
	and the scaled time
	$\frac{2\pi\tau}{T}\to \tau$,   
	it follows that the last expressions become
	$$
	T^{tt}(k,\omega) = \frac{T_s}{(1-v^2)}\frac{L}{\pi^2}\int^{\pi}_{0} d\tau \int^{2\pi}_{0}  d\sigma ( I^* e^{2i (\sigma + 2\tau)}+I e^{-2i(\sigma + 2\tau)}+J)\;e^{-i2n\alpha \sin\sigma}e^{-i2n\beta \cos\sigma}e^{i2 n \tau},
	$$
	$$
	T^{xx}(k,\omega) = -\frac{T_s}{(1-v^2)}\frac{L}{\pi^2}\int^{\pi}_{0} d\tau \int^{2\pi}_{0}  d\sigma \cos^2\sigma\;( I^* e^{2i (\sigma + 2\tau)}+I e^{-2i(\sigma + 2\tau)}+J)\;e^{-i2n\alpha \sin\sigma}e^{-i2n\beta \cos\sigma}e^{i2 n \tau},
	$$
	$$
	T^{yy}(k,\omega) = -\frac{T_s}{(1-v^2)}\frac{L}{\pi^2}\int^{\pi}_{0} d\tau \int^{2\pi}_{0}  d\sigma \sin^2\sigma\;( I^* e^{2i (\sigma + 2\tau)}+I e^{-2i(\sigma + 2\tau)}+J)\; e^{-i2n\alpha \sin\sigma}e^{-i2n\beta \cos\sigma}e^{i2 n \tau},
	$$
	$$
	T^{xy}(k,\omega) = \frac{T_s}{(1-v^2)}\frac{L}{\pi^2}\int^{\pi}_{0} d\tau \int^{2\pi}_{0}  d\sigma \frac{1}{2}\sin2\sigma\;( I^* e^{2i (\sigma + 2\tau)}+I e^{-2i(\sigma + 2\tau)}+J)\;e^{-i2n\alpha \sin\sigma}e^{-i2n\beta \cos\sigma}e^{i2 n \tau}.
	$$
	In the last formulas, the quantities $I$ and $J$
		$$
	I = (\frac{1}{6}\omega_1^2+\frac{1}{9}\omega_2^2\kappa)[\frac{1}{4}(U^2-V^2)+\frac{1}{2}iW],
	$$
	\begin{equation}\label{dame}
		J =  \frac{1}{2}(U^2+V^2)(\frac{1}{6}\omega_1^2+\frac{1}{9}\omega_2^2\kappa)+a_c^2\omega_1^2 - 2,
	\end{equation}
	have been introduced. At this point it is useful to consider  the integral in $\tau$ by employing the formula
	\begin{equation}
		\int^\pi_0 \exp\{2i(n\pm m)\tau\}d\tau =\frac{\sin[(n\pm m)2\pi]}{n\pm m},
	\end{equation}
	for the particular cases $m = 0, 2$. With the help of the last expression, it is arrived to
	$$
	T^{tt}(k,\omega) = \frac{T_s}{(1-v^2)}\frac{L}{\pi^2} \int^{2\pi}_{0}  d\sigma \bigg[ \frac{I^*\sin(2\pi n)}{n + 2}e^{2i \sigma}+ \frac{I\sin(2\pi n)}{n-2}e^{-2i\sigma}
	+\frac{J\sin(2\pi n)}{n}\bigg]\;e^{-i2n\alpha \sin\sigma}e^{-i2n\beta \cos\sigma},
	$$
	$$
	T^{xx}(k,\omega) = -\frac{T_s}{(1-v^2)}\frac{L}{\pi^2} \int^{2\pi}_{0}  d\sigma \cos^2\sigma\;\bigg[ \frac{I^*\sin(2\pi n)}{n + 2}e^{2i \sigma}+ \frac{I\sin(2\pi n)}{n-2}e^{-2i\sigma}
	$$
	$$
	+\frac{J\sin(2\pi n)}{n}\bigg]\;e^{-i2n\alpha \sin\sigma}e^{-i2n\beta \cos\sigma},$$
	$$         T^{yy}(k,\omega) = -\frac{T_s}{(1-v^2)}\frac{L}{\pi^2} \int^{2\pi}_{0}  d\sigma \sin^2\sigma\;\bigg[ \frac{I^*\sin(2\pi n)}{n + 2}e^{2i \sigma}+ \frac{I\sin(2\pi n)}{n-2}e^{-2i\sigma}
	$$
	$$
	+\frac{J\sin(2\pi n)}{n}\bigg]\; e^{-i2n\alpha \sin\sigma}e^{-i2n\beta \cos\sigma},$$
	$$
	T^{xy}(k,\omega) = \frac{T_s}{2(1-v^2)}\frac{L}{\pi^2}\int^{2\pi}_{0}  d\sigma\sin2\sigma\;\bigg[ \frac{I^*\sin(2\pi n)}{n + 2}e^{2i \sigma}+ \frac{I\sin(2\pi n)}{n-2}e^{-2i\sigma}
	$$
	$$
	+\frac{J\sin(2\pi n)}{n}\bigg]\;e^{-i2n\alpha \sin\sigma}e^{-i2n\beta \cos\sigma}.
	$$
	As $n>0$, the first term in the square brackets always will be zero.  To streamline the computation,  the freedom in adjusting the constants $U$, $V$, and $W$ within the solution (\ref{General_Sol}) can be employed. One may impose $J=0$ with the imposition of the following constraint
	\begin{equation}
		U^2+V^2 = 2[2 - a_c^2\omega_1^2](\frac{1}{6}\omega_1^2+\frac{1}{9}\omega_2^2\kappa)^{-1},
	\end{equation}
	this follows from \eqref{dame}. By employing this choice it is found explicitly that
	$$
	T^{tt}(k,\omega) = \frac{T_s}{(1-v^2)}\frac{L}{\pi^2}\frac{I\sin(2\pi n)}{n-2} \int^{2\pi}_{0}  d\sigma e^{-2i\sigma}\;e^{-i2n\alpha \sin\sigma}e^{-i2n\beta \cos\sigma},$$
	$$
	T^{xx}(k,\omega) = -\frac{T_s}{(1-v^2)}\frac{L}{\pi^2}\frac{I\sin(2\pi n)}{n-2} \int^{2\pi}_{0}  d\sigma \cos^2\sigma\;e^{-2i\sigma}e^{-i2n\alpha \sin\sigma}e^{-i2n\beta \cos\sigma},$$
	$$
	T^{yy}(k,\omega) = -\frac{T_s}{(1-v^2)}\frac{L}{\pi^2}\frac{I\sin(2\pi n)}{n-2} \int^{2\pi}_{0}  d\sigma \sin^2\sigma\; e^{-2i\sigma}e^{-i2n\alpha \sin\sigma}e^{-i2n\beta \cos\sigma},$$
	$$
	T^{xy}(k,\omega) = \frac{T_s}{(1-v^2)}\frac{L}{\pi^2}\frac{I\sin(2\pi n)}{n-2}\int^{2\pi}_{0}  d\sigma \frac{1}{2}\sin2\sigma\;e^{-2i\sigma}e^{-i2n\alpha \sin\sigma}e^{-i2n\beta \cos\sigma}.
	$$
	While this specific choice of constants may seem somewhat restrictive, it proves adequate for illustrating an example of a static and highly radiating object, which is one of the objectives of the present work. By further taking into account that
	\begin{equation}
		\cos^2\sigma = \frac{1}{4}(e^{2i\sigma} + e^{-2i\sigma} + 2),\qquad
		\sin^2\sigma= -\frac{1}{4}(e^{2i\sigma} + e^{-2i\sigma} - 2),\qquad
		\frac{1}{2}\sin2\sigma = \frac{1}{4i}(e^{2i\sigma} - e^{-2i\sigma}),
	\end{equation}
	it is seen that the integrals  to be solved are of the following type
	\begin{equation}
		\mathcal{I}=\int_0^{2\pi}d\sigma e^{-ir\sigma}e^{-2in\alpha\sin\sigma}e^{-2in\beta\cos\sigma},
	\end{equation}
	where $r$ might be $0,2,4$.
	By setting $\alpha$ and $\beta$ for their angular coordinates, which is to say that $\alpha = \cos\phi \sin\theta,\; \beta=\sin\phi \sin\theta$, it is arrived to
	\begin{equation}
		\mathcal{I}=\int_0^{2\pi}d\sigma e^{-ir\sigma}e^{-2in\sin\theta\sin(\sigma+\phi)}.
	\end{equation}
	The definition of the Bessel functions
	\begin{equation}
		2\pi J_k(x) = (-1)^k\int_0^{2\pi} d\sigma e^{i(k\sigma+x\sin\sigma)}
	\end{equation}
	leads to
	$$
	\mathcal{I}=e^{2ir\phi}\int_\phi^{2\pi+\phi}d\sigma e^{-2ir\sigma}e^{-2in\sin\theta\sin\sigma}= 2\pi e^{2ir\phi} (-1)^{2r} J_{-2r}(-2n\sin\theta)= 2\pi e^{2ir\phi} J_{2r}(2n\sin\theta).
	$$
	Whit the help of the last expression, the non trivial components of the energy tensor become
	$$
	T^{tt}(k,\omega) = \frac{T_s}{(1-v^2)}\frac{L}{\pi^2}\frac{I\sin(2\pi n)}{n-2} 2\pi e^{4i\phi} J_{4}(2n\sin\theta),
	$$
	$$
	T^{xx}(k,\omega) = -\frac{T_s}{(1-v^2)}\frac{L}{\pi^2}\frac{I\sin(2\pi n)}{n-2} \frac{\pi}{2} \bigg[ J_{0}(2n\sin\theta) + e^{8i\phi} J_{8}(2n\sin\theta) + 2 e^{4i\phi} J_{4}(2n\sin\theta)\bigg],
	$$
	$$
	T^{yy}(k,\omega) = \frac{T_s}{(1-v^2)}\frac{L}{\pi^2}\frac{I\sin(2\pi n)}{n-2}  \frac{\pi}{2} \bigg[ J_{0}(2n\sin\theta) + e^{8i\phi} J_{8}(2n\sin\theta) - 2 e^{4i\phi} J_{4}(2n\sin\theta)\bigg],$$
	$$
	T^{xy}(k,\omega) = \frac{T_s}{(1-v^2)}\frac{L}{\pi^2}\frac{I\sin(2\pi n)}{n-2} \frac{\pi}{2i} \bigg[ J_{0}(2n\sin\theta) - e^{8i\phi} J_{8}(2n\sin\theta)\bigg].
	$$
	The last expression was derived for generic $n$. However,  as mentioned above, the case of interest is  $n=2$. In this situation
	$$
	T^{tt}(k,\omega_2) = \frac{T_s}{(1-v^2)}4L\,I e^{4i\phi} J_{4}(4\sin\theta),
	$$
	$$
	T^{xx}(k,\omega_2) = -\frac{T_s}{(1-v^2)}LI  \bigg[ J_{0}(4\sin\theta) + e^{8i\phi} J_{8}(4\sin\theta) + 2 e^{4i\phi} J_{4}(4\sin\theta)\bigg],
	$$
	$$
	T^{yy}(k,\omega_2) = \frac{T_s}{(1-v^2)}LI \bigg[ J_{0}(4\sin\theta) + e^{8i\phi} J_{8}(4\sin\theta) - 2 e^{4i\phi} J_{4}(4\sin\theta)\bigg],
	$$
	\begin{equation}\label{dime}
		T^{xy}(k,\omega_2) = -\frac{T_s}{(1-v^2)}LI \bigg[ J_{0}(4\sin\theta) - e^{8i\phi} J_{8}(4\sin\theta)\bigg].
	\end{equation}
	On the other hand, the power radiated 
	\begin{equation}
		\frac{dP}{d\Omega} = \frac{G\omega_2^2}{\pi}[ T^*_{\mu\nu}(\omega_2,k)T^{\mu\nu}(\omega_2,k)-\frac{1}{2}|T^{\nu}_{\nu}(\omega_2,k)|^2],
	\end{equation}
	may be easily expressed as
	\begin{equation}
		\frac{dP}{d\Omega} = \frac{G\omega_2^2}{\pi}[ |T^{tt}|^2+|T^{xx}|^2+|T^{yy}|^2+2|T^{xy}|^2-\frac{1}{2}|-T^{tt}+T^{xx}+ T^{yy}|^2],
	\end{equation}
	or equivalently, as
	\begin{equation}\label{eqi}
		\frac{dP}{d\Omega} = \frac{G\omega_2^2}{\pi}\bigg[ \frac{1}{2}|T^{tt}|^2+\frac{1}{2}|T^{xx}|^2+\frac{1}{2}|T^{yy}|^2+2|T^{xy}|^2
		-\text{Re} ( T^{*tt}T^{xx}+T^{*tt}T^{yy}-T^{*xx}T^{yy})\bigg].
	\end{equation}
	The intensity as a function of $\theta$ of this power radiated is shown in the figure below.
	\begin{figure}
		\includegraphics[width=\linewidth]{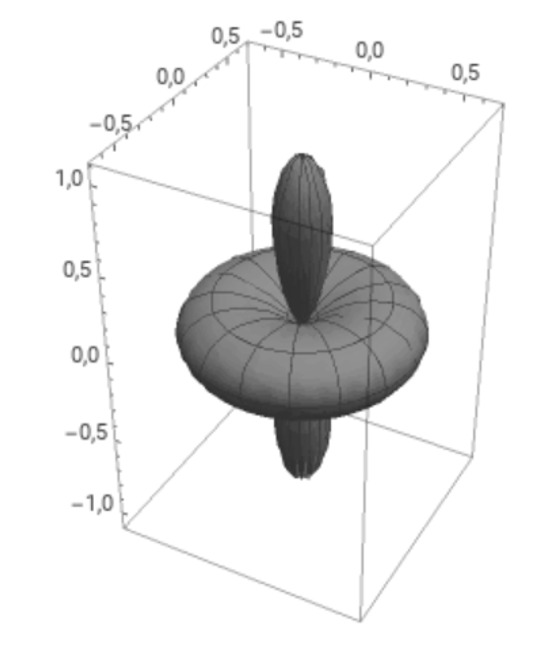}
		\caption{Power radiated intensity. The vertical is the z axis}
		\label{fig:boat1}
	\end{figure}
	The equation  \eqref{dime}  leads in particular to
	$$
	|T^{tt}|^2 = \frac{T_s^2 L^2 |I|^2}{(1-v^2)^2}\, 16 J^2_{4}(4\sin\theta),$$
	$$
	|T^{xx}|^2 = \frac{T_s^2 L^2 }{(1-v^2)^2}|I|^2 \bigg\{  J^2_{0} + J^2_{8} + 4 J^2_{4}
	+ 4J_0J_4\cos4\phi+2J_0J_8\cos8\phi+4J_4J_8\cos4\phi \bigg\}(4\sin\theta),$$
	$$
	|T^{yy}|^2 = \frac{T_s^2 L^2|I|^2}{(1-v^2)^2} \bigg\{  J^2_{0} + J^2_{8} + 4 J^2_{4}
	- 4J_0J_4\cos4\phi+2J_0J_8\cos8\phi-4J_4J_8\cos4\phi \bigg\}(4\sin\theta),
	$$
	$$
	|T^{xy}|^2 = \frac{T_s^2 L^2 |I|^2}{(1-v^2)^2} \bigg\{  J^2_{0} + J^2_{8}- 2J_0J_8\cos8\phi \bigg\}(4\sin\theta),
	$$
	$$
	\text{Re}T^{*tt}T^{xx} = -\frac{T_s^2 L^2 |I|^2}{(1-v^2)^2} 4\bigg\{ 2J^2_{4} +J_0J_4\cos4\phi +J_4J_8\cos8\phi\bigg\}(4\sin\theta),
	$$
	$$
	\text{Re}T^{*tt}T^{yy} = -\frac{T_s^2 L^2 |I|^2}{(1-v^2)^2} 4\bigg\{ 2J^2_{4} -J_0J_4\cos4\phi -J_4J_8\cos8\phi\bigg\}(4\sin\theta),
	$$
	$$
	\text{Re}T^{*xx}T^{yy} = -\frac{T_s^2 L^2 |I|^2}{(1-v^2)^2} \bigg\{ J_0^2+J_8^2 - 4J_4^2+2J_0J_8\cos8\phi\bigg\}(4\sin\theta).
	$$
	With these formulas at hand, by integrating \eqref{eqi} along the angles $\phi$ and $\theta$ it is arrived to the following power radiated
	\begin{equation}\label{disco}
		P=\frac{32\pi G T_s^2 |I|^2}{(1-v^2)^2}\int_0^{\pi} (4J_0^2 + 4 J_8^2 + 24J_4^2)(4\sin\theta) \sin\theta\; d\theta.
	\end{equation}

	The main frequency considered here is $\omega_2$. Under this approximation, it has been found that the last integral in independent on the size $L$ of the vortex ring, a result that has been derived for a circular ring. Therefore, the power radiated spectrum depends on the shape but not on the size
	of the loop. Concerning the finite factor
	\be\lb{replas}
	\delta_R=\int_0^{\pi} (4J_0^2 + 4 J_8^2 + 24J_4^2)(4\sin\theta) \sin\theta\; d\theta,
	\ee
	the authors were unable to find an explicit expression for this integral, however by use of MATHEMATICA its 
	estimated value is 
	$$
	\delta_R\sim 5,7.
	$$
	In these terms, the expression \eqref{disco} takes the form
	\begin{equation}\label{factors}
	P_{\text{non abelian}}=\gamma_l T_s^2 G,
	\end{equation}
	in which the loop factor becomes $\gamma_l\sim 500|I|^2$.

	\section{Summary}
	In the present letter, the possibility that a non abelian, almost static, semi local vortex may emit a considerable amount of gravitational radiation was studied. The main formula obtained is \eqref{factors}. This formula may be compared with the gravitational radiation of translational modes in an abelian vortex. First of all, it should be remarked that the factor $T_s^2 G$ appearing in that expression also appears when translational modes are excited \cite{galaxy}-\cite{kibble}. In fact typically in those situations the power radiated is of the form 
	$$
	P_{\text{abelian}}=\gamma_{abelian} T_s^2 G,
	$$
	where $\gamma_{abelian}$ is a loop factor depending on the geometry, typically taking
	values between $1-100$. By comparing this expression with \eqref{factors}, it follows that for the present non abelian vortex the effective loop factor will be
	$$
	\gamma_l\sim \frac{16 |I|^2\delta_R}{ (1-v^2)^2}.
	$$
	
	Although it may appear that this factor can be very large for ultra relativistic excitations $v\sim 1$, for such large velocities the use of the present formula for radiated power may be dubious. Note that this factor, as discussed below \eqref{factors}, is of order $\gamma_l\sim 500|I|^2$, and can overcome the value $\gamma_{abelian}=100$ for moderate values of $|I|^2$. This suggest that internal modes may generate a considerable amount of gravitational radiation. In fact, the quantity $I$ was defined in \eqref{dame}, and an inspection of this formula shows that it depends on the maximum radius $\rho_m$ of motion and the frequencies $\omega_i$. This last quantity describes the  rotation inside one of the $SU(2)$ parts of the conifold geometry. It may be expected that $|I|^2 \sim F_2(\rho_m \Omega_\tau, \rho_m \Omega_\sigma)$. In fact, \eqref{dame} shows that  it is a quartic expression in those variables. Therefore, the loop factor for this static vortex looks like 
	$$
	\gamma_l\sim \frac{16 \rho_m^4 \omega^4\delta_R}{ (1-v^2)^2},
	$$
	with $\omega$ some average of the rotation frequencies in the $SU(2)$ component of the conifold. The use of the Manton approximation 
	does not forbid at first sight to overcome the value $\gamma_l=100$ in several particular situations. In the present case, there is no translational contribution, as the loop motion was neglected and all the calculated gravitational spectrum is entirely due to the internal modes. This is a clear difference with respect to abelian vortices since the last ones do not radiate when they are static. Here, instead, it does.

	In a general picture, it is expected that a  typical non abelian vortex with a non compact space of internal excitations may move as well. The suggestion of the present work is that the radiation due to the internal modes may predominate in several cases. Therefore, this excess of of gravitational radiation or a very large correction of the loop factor may be considered as a possible signature of non abelianity.  For this reasons, we suggest that the study of radiation channels for non abelian vortices with non compact internal moduli spaces is a topic that deserve further attention.

\section*{Acknowledgments}
O.P.S supported by CONICET, Argentina and  by the Grant PICT 2020-02181.

	\end{document}